\documentclass[aps,prb,twocolumn,superscriptaddress]{revtex4-2}
\usepackage[usenames,dvipsnames]{xcolor}
\usepackage{amssymb}
\usepackage{graphicx}
\usepackage{amsmath}
\usepackage{bbm}
\usepackage[bookmarks=true,colorlinks,citecolor=blue,urlcolor=blue]{hyperref}
\usepackage{braket}
\usepackage{babel}
\usepackage{blindtext}
\usepackage[compat=1.1.0]{tikz-feynman}
\usepackage[normalem]{ulem}
\usepackage{physics}
\usepackage{mathrsfs}
\usepackage{footmisc}
\usepackage{booktabs}

\usepackage{mathtools}

\usepackage{dsfont}

\definecolor{mypurp}{rgb}{0.35, 0, 0.7}

\usepackage{amsthm}
\usepackage{txfonts}

\usepackage{soul}

\theoremstyle{definition}

\begin{document}

\title{Bridging Rokhsar-Kivelson Type and Generic Quantum Phase Transitions\\ via Thermofield Double States}
\newcommand{\tsinghua}[0]{State Key Laboratory of Low-Dimensional Quantum Physics and Department of Physics, Tsinghua University, Beijing 100084, China.}

\author{Wen-Tao Xu}
\affiliation{\tsinghua}
\affiliation{Technical University of Munich, TUM School of Natural Sciences, Physics Department, 85748 Garching, Germany}
\affiliation{Munich Center for Quantum Science and Technology (MCQST), Schellingstr. 4, 80799 M{\"u}nchen, Germany}

\author{Rui-Zhen Huang}
\affiliation{Department of Physics and Astronomy, University of Ghent, Belgium}

\author{Guang-Ming Zhang}
\affiliation{\tsinghua}
\affiliation{Frontier Science Center for Quantum Information, Beijing 100084, China.}
\date{\today}

\begin{abstract}
The formalism of the Rokhsar-Kivelson (RK) model has been frequently used to study topological phase transitions in 2D in terms of the deformed wavefunctions, which are RK-type wavefunctions. A key drawback of the deformed wavefunctions is that the obtained quantum critical points are RK-type, in the sense that the equal-time correlation functions are described by 2D conformal field theories (CFTs). The generic Lorentz invariant quantum critical points described by (2+1)D CFTs can not be obtained from the deformed wavefunctions. To address this issue, we generalize the deformed wavefunction approach to the deformed thermofield double (TFD) state methodology. Through this extension, we can effectively reconstruct the absent temporal dimension at the RK-type quantum critical point. We construct deformed TFD states for a (1+1)D quantum phase transition from a symmetry-protected topological phase to a symmetry-breaking phase, and for generic (2+1)D topological phase transitions from a $\mathbb{Z}_2$ topologically ordered phase to a trivial paramagnetic phase.
\end{abstract}

\maketitle

\section{Introduction}
In recent years, topological phases of matter and their quantum phase transitions have attracted enormous research interest. Particularly, there has been a very active development of tensor network states (TNS)~\cite{Verstraete-Cirac-2004,Nishino-2001,Nishino-2004} to study topological phases of matter~\cite{TNS_review_RMP_2021,schuch_peps_2010,BULTINCK_2017}, because ground state entanglement entropy of gapped quantum systems usually satisfies the area law~\cite{Wolf-Cirac-2008,Eisert-rmp-2010}, which is the fundamental reason that TNS can efficiently approximate (non-chiral) topological state of gapped systems.

Despite the great success in classifying gapped systems, in general, TNS can not efficiently approximate the ground state of a quantum critical point or gapless phase. However, counter-intuitively, there exists a class of critical models in (2+1)D called the Rokhsar-Kivelson (RK) type model~\cite{RK_1998}, whose ground states dubbed RK-type wavefunction can be exactly expressed in terms of 2D TNS~\cite{Verstraete_2006}, which is also called projected entangled pair states (PEPS). The effective field theories describing the RK-type models are not Lorentz invariant because their low-energy excitations have a dispersion $\omega \sim k^z$ with a dynamical critical exponent $z>1$~\cite{ardonne_2004,castelnovo-2005,Isakov-2011}. The equal-time correlation functions of RK-type wavefunctions are usually described by 2D conformal field theories (CFTs)~\cite{ardonne_2004,castelnovo-2005,Isakov-2011}.

RK-type wavefunctions frequently appear when studying topological phase transitions via an approach called deformed wavefunctions. Traditionally, to study topological phase transitions, one constructs a one-parameter family of Hamiltonians by interpolating fixed point Hamiltonians of the topological phase and the trivial phase and then solves the ground states, which is usually very difficult for $(2+1)$D and higher dimensional quantum systems. Another simple yet systematic approach is the deformed wavefunction approach. Instead of interpolating fixed point Hamiltonians, one can construct a one-parameter family of wavefunctions by deforming a fixed point wavefunction of the topological phase towards the fixed point product state of the trivial phase, such that a topological phase transition is included in the one-parameter family of deformed wacefunctions.  Interestingly, the norm of a $d$-dimensional deformed wavefunction can usually be interpreted as a partition function of a known $d$-dimensional classical model. The order parameters of the classical model can be used to characterize the topological phase transitions~\cite{zhu_gapless_2019}. The deformed wavefunction approach has been used to study various topological phase transitions, including the phase transitions between symmetry-protected topological (SPT) states~\cite{Wolf-Cirac-2008,Huang_wei_2016,Skeleton_2021}, topological phase transitions of Abelian topological states~\cite{Simon_Trebst_2007,zhu_gapless_2019,haegeman_shadows_2015,Iqbal_2017,Iqbal_2018,xu_zhang_2018,Qi_zhang_2020} and non-Abelian topological states~\cite{condensation_deriven_2017,Xu_fib_2020,xu_2021,xu_2022}, topological phase transitions between symmetry enriched topological states~\cite{xu_2023}, as well as phase transitions of fracton topological states~\cite{zhu_2023}.

However, the deformed wavefunction approach has a key drawback. Only RK-type quantum critical points, which is usually called conformal quantum critical point in literatures~\cite{ardonne_2004,Simons_Trebst_2009,Isakov-2011}, can be obtained from the deformed wavefunction, and generic quantum critical points with the Lorentz invariance are missing.  At a generic quantum critical point with Lorentz invariance, one has to increase the bond dimensions of the PEPS and perform finite-entanglement scaling~\cite{Pollmann_2009,Corboz-prx-2018,Lauchli-prx-2018,Bram_2019,Bram_2022}. However, the deformed wavefunctions are usually TNS with a fixed bond dimension, and the entanglement entropy satisfies the area law exactly, which leaves no room for finite entanglement scaling. At the RK-type quantum critical point, the correlations are strongly suppressed in the temporal direction compared to the spatial one, so dimensionality reduction happens~\cite{Simons_Trebst_2009}. To obtain generic quantum critical points with emergent Lorentz invariance, one needs to modify the deformed wavefunction approach by taking strong temporal fluctuations or correlations into account to reconstruct the missing temporal dimension.

In this paper, we generalize the deformed wavefunction approach using the so-called thermofield double (TFD) states~\cite{THERMO_1996}, which are defined in an enlarged Hilbert space $\mathscr{H}_{P}\otimes \mathscr{H}_{F}$ consisting of the physical space $\mathscr{H}_{P}$ and a fictitious space $\mathscr{H}_{F}$. The fictitious space accounts for the missing temporal fluctuations or correlations in the conventional deformed wavefunction. The TFD state can be viewed as a purification of a density operator in thermal equilibrium~\cite{Frank_V_purification_1D_2004}, and the thermal fluctuations are encoded in the form of quantum entanglement between the physical and fictitious parts. We construct a general framework of the deformed TFD states, from which we can obtain generic topological phase transitions with emergent Lorentz invariance. Importantly, when the physical and fictitious degrees of freedom are disentangled, the deformed TFD approach naturally reduces the conventional deformed wavefunction approach,  so we can understand how an RK-type quantum critical point and a generic Lorentz invariant quantum critical point are connected through the deformed TFD state approach.

The rest of the paper is organized as follows. In Sec.~\ref{framework}, we introduce the deformed TFD state framework, generalizing the deformed wavefunctions. In Sec.~\ref{1_plus_1_d_TFD}, we show a one-parameter family of $(1+1)$D deformed TFD states describing a generic topological phase transition with Lorentz invariance between an SPT phase and a symmetry-breaking phase. In Sec.~\ref{2_plus_1_d_TFD}, we construct a two-parameter family of $(2+1)$D deformed  TFD states for Lorentz invariant topological phase transitions between a $\mathbb{Z}_2$ topologically ordered phase and a trivial paramagnetic phase~\cite{kitaev_toric_code_2003,Vidal2009,youjin2012}. Finally, we give a summary and outlook in Sec.~\ref{sec_summary}.

\section{General framework for the deformed TFD states}\label{framework}
In this section, we introduce the deformed TFD states for generic topological phase transitions. First, we can easily construct the TFD states for fixed points of a topological phase and a trivial phase. Then, we introduce a one-parameter family of deformed TFD states by interpolating between different fixed-point TFD states. A reduced TFD density operator can be obtained by tracing the fictitious degrees of freedom of a deformed TFD state. It can be interpreted as the transfer matrix of a (isotropic) classical model  whose dimensionality is one plus the spatial dimensionality of the deformed TFD state. Therefore generic continuous topological phase transitions with emergent Lorentz invariance can be obtained from the deformed TFD states. More interestingly, the order parameter of the classical model can signal this topological phase transition.

Before introducing the deformed TFD states, we first review the deformed wavefunction approach and its connection to classical models. While here we focus on (2+1)D, it's worth noting that the framework is versatile and applicable across various dimensions.
Given a 2D local classical Hamiltonian $E(\sigma
_{1},\cdots ,\sigma _{n})$ with spin variables $\{\sigma _{i}\}$, a
RK wavefunction with a parameter $\beta $ can be defined as%
\begin{equation}
|\tilde{\psi}(\beta )\rangle =\sum_{\pmb{\sigma}}\exp\left[\frac{-\beta E(\pmb{\sigma})}{2}\right]|%
\pmb{\sigma}\rangle,  \label{RK}
\end{equation}%
where $|\pmb{\sigma}\rangle =|\sigma _{1},\cdots ,\sigma _{n}\rangle $ and $%
\{|\pmb{\sigma}\rangle \}$ is a set of complete and orthonormal basis. It is easy to express the RK wavefunction in terms of an exact PEPS with a finite bond dimension as long as the classical Hamiltonian $E(\sigma_1\cdots\sigma_n)$ only involves local interactions, and a parent Hamiltonian of the RK wavefunction, called RK Hamiltonian, can be constructed~\cite{Verstraete_2006}. The
wavefunction norm is precisely a partition function of a
classical model with $\beta $ as the inverse temperature:
\begin{equation}\label{RK_partition_function}
Z(\beta )=\langle \tilde{\psi} (\beta )|\tilde{\psi} (\beta )\rangle =\sum_{\pmb{\sigma}}\exp
[-\beta E(\pmb{\sigma})].
\end{equation}%
With such a quantum-classical mapping, all equal-time correlation functions of
the RK wavefunction become the correlation functions of the
classical partition function. So, the static properties of the RK Hamiltonian near and at the quantum critical point are controlled by the 2D CFT describing the critical point of the 2D classical model.

The deformed wavefunctions are very similar to the RK wavefunctions. Consider a fixed point Hamiltonian $\hat{H}_0$ of a topological phase and a fixed point Hamiltonian $\hat{H}_1$ of a trivial phase, we can deform a fixed point topological state towards a fixed point of a trivial state:
\begin{equation}\label{deformed_TNS}
\ket{\psi(\alpha_1)}=\exp(-\frac{\alpha_1 H_1}{2})\ket{\psi_0},
\end{equation}
where $\ket{\psi_0}$ [ $\ket{\psi_1}=\ket{\psi(+\infty)}$] is a ground state of $H_{0}$ [$H_{1}$] and $\alpha_1$ is a tuning parameter. Since  $\ket{\psi_0}$ as a fixed point wavefunction of a topological phase can be exactly expressed as a PEPS~\cite{Gu_string_net_PEPS_2009,G_vidal_string_net_PEPS_2009}, and  $H_1$ as a fixed point Hamiltonian only consists of local commuting terms, $\ket{\psi(\alpha_1)}$ can be exactly expressed as a one-parameter family of PEPS. Notice that the deformed wavefunction $\ket{\psi(\alpha_1)}$ in Eq.~\eqref{deformed_TNS} and the RK wavefunction in Eq.~\eqref{RK} are very similar, both of them can be expressed a constant bond dimension PEPS and mapped to 2D classical models, so the deformed wavefunction $\ket{\psi(\alpha_1)}$ in Eq.~\eqref{deformed_TNS} is called RK-type wavefunction. One essential difference between the RK wavefunction in Eq.~\eqref{RK} and the deformed wavefunction in Eq.~\eqref{deformed_TNS} is that the corresponding classical partition function of later could have negative Blotzmann weights~\cite{Xu_fib_2020,Qi_zhang_2020}.

To describe a generic Lorentz invariant (2+1)D quantum critical point, we propose the deformed TFD states approach. Given a 2D quantum Hamiltonian $H$ defined in the Hilbert space spanned by the orthonormal basis $\{|\pmb{\sigma}\rangle \}$, a TFD state describing the quantum system at a finite temperature $1/\beta$ is
\begin{equation}
|\tilde{\Psi}(\beta )\rangle =\sum_{\pmb{\sigma}}\left[\exp\left(- \frac{\beta H}%
{2}\right)|\pmb{\sigma}\rangle _{P}\right] \otimes |\pmb{\sigma}\rangle _{F},
\label{TFD}
\end{equation}%
where the enlarged Hilbert space $\mathscr{H}_{P}\otimes \mathscr{H}_{F}$
consists of a physical space $\mathscr{H}_{P}$ and a fictitious one $%
\mathscr{H}_{F}$, and $\exp\left(-\beta H/2\right)$ acts only on $\mathscr{H}_{P}$ . It is always sufficient to choose $\mathscr{H}_{F}$ to
be identical to $\mathscr{H}_{P}$, ``doubling'' original physical degrees of freedom. Unlike the RK-type wavefunctions, usually, one can not exactly express a TFD state as a PEPS~\cite{Approximating_Gibbs_2015}. Tracing out the degrees of freedom in the fictitious space yields a
reduced density matrix of the TFD state:
\begin{equation}
\rho=\text{tr}_{F}|\tilde{\Psi}(\beta )\rangle \langle \tilde{\Psi}(\beta )|=\exp\left(-\beta \hat{H}\right),
\end{equation}%
which is the Gibbs density matrix. The information of thermal fluctuations is encoded in the form of the quantum entanglement between
the physical and fictitious spaces. The norm of the TFD state corresponds
to the partition function of the 2D quantum Hamiltonian:
\begin{equation}
\mathcal{Z}(\beta )=\langle \tilde{\Psi}(\beta )|\tilde{\Psi}(\beta
)\rangle =\Tr\exp\left(-\beta H\right).
\end{equation}%
Moreover, in the
limit of $\beta \rightarrow +\infty $, if the system is gapped and the ground state is non-degenerate, $|\tilde{\Psi}(+\infty)\rangle =|\psi\rangle _{P}\otimes |\psi^{\ast
}\rangle _{F}$, where $\ket{\psi}$ is a ground state of $H$.  However, it is essential for certain gapless
systems, i.e., a Lorentz invariant critical point whose partition function is not only determined by the ground state, to consider the entanglement between the physical and fictitious spaces of a TFD state.

Inspired by a variational ansatz in Ref.~\cite{Bridging2017}, we propose the following deformed TFD state:
\begin{equation}
|\Psi(\alpha _{0},\alpha_{1})\rangle =\sum_{\pmb{\sigma}}
\left[\exp\left(-\frac{\alpha_0 H_0}{2}\right)\exp\left(-\frac{\alpha_1 H_1}{2}\right)
|\pmb{\sigma}\rangle _{P}\right]\otimes |\pmb{\sigma}\rangle _{F},
\label{Main}
\end{equation}%
where $(\alpha_{0},\alpha_{1})$ are two turning parameters, $H_0$ and $H_1$ are fixed point Hamiltonian of a  topological phase and a trivial phase. Unlike the original TFD state shown in Eq.~\eqref{TFD}, the deformed TFD state in Eq.~\eqref{Main} can be exactly expressed as a PEPS as long as $H_0$ and  $H_1$ consists of local commuting terms, separately. Notice that $\ket{\Psi(+\infty,0)}=\ket{\psi_0}_P\otimes \ket{\psi^*_0}_F,\quad  \ket{\Psi(0,+\infty)}=\ket{\psi_1}_P\otimes \ket{\psi^*_1}_F$ are chosen as fixed-point TFD states representing different phases, so the one-parameter family of deformed TFD states in Eq.~\eqref{Main} interpolate between the fixed points TFD states of the topological phases and the trivial phases. Furthermore, because
\begin{equation}\label{disentangle}
    \ket{\Psi(+\infty,\alpha_1)}\propto\ket{\psi_0}_P\otimes \ket{\psi^*(\alpha_1)}_F,
\end{equation}
the deformed wavefunction in Eq.~\eqref{deformed_TNS} is included in the deformed TFD state in Eq.~\eqref{Main}.

After tracing the fictitious degrees of
freedom of deformed TFD state in Eq.~\eqref{Main}, we obtain a reduced TFD density matrix:
\begin{equation}
\rho =\text{tr}_{F}|\Psi\rangle \langle \Psi|=\exp\left(-\frac{\alpha _{0}%
H_0}{2}\right)\exp\left(-\alpha _{1}H_1\right)\exp\left(-\frac{\alpha_{0}H_0}{2}\right),\notag
\end{equation}%
where an entanglement temperature $T_E=1$ implicitly included $\rho$~\cite{Li_haldane_2008,How_universal_2014}. To approach quantum phases and phase transitions at $T_E=0$, we consider $\rho^N$ with $N\in\mathbb{Z}_{+}$ to reduce the entanglement temperature $T_E$ from $1$ to $1/N$. Notice that the purification of $\rho^N$ is another TFD state whose bond dimension is larger than that of $\ket{\Psi}$, so the bond dimension is effectively increasing by reducing the entanglement temperature $T_E$. Then, a partition function $\mathcal{Z}_{N}=$tr$\rho ^{N}$ can be obtained in the Euclidian spacetime. Because the deformed TFD state in Eq.~\eqref{Main} can be exactly expressed as a 2D PEPS, the partition function $\mathcal{Z}_{N}$ can be exactly expressed as a 3D tensor network when $N\rightarrow\infty$. In practice, we can prepare a deformed TFD state with a finite system size $L\times L$, which can be efficiently simulated using Monte Carlo methods. To approach (2+1)D quantum criticality, it suffices to consider a finite $N\propto L$, as long as the system enters the scaling regime. In contrast, if  $N$ is fixed and $L\gg N$, we can obtain the (2+0)D criticality from the deformed TFD state. 

So, via generalizing the deformed wavefunctions to the deformed TFD states, we can construct the missing temporal dimension at the RK point. There are two essential steps from RK-type quantum phase transitions to the generic quantum phase transitions. 
First, from Eq.~\eqref{disentangle}, the deformed wavefunctions are the deformed TFD states whose physical and fictitious spaces are disentangled, so the entanglement between them is  essential to go beyond the RK-type deformed wavefunctions. Second, via gradually reducing the entanglement temperature $T_E=1/N$, one could get a crossover from an RK-type quantum critical point with dimensionality reduction to a generic quantum critical point with an expected dimensionality. 

\begin{figure}[tbp]
\centering
\includegraphics{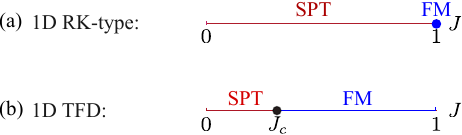}
\caption{(a) The phase diagram of the 1D RK-type deformed SPT state $\ket{\psi(J)}$ shown in Eq.~\eqref{RK_SPT}. $\ket{\psi(J<1)}$ is a SPT state, and $\ket{\psi(J=1)}$ a ferromagnetic state. (b) The phase diagram of the 1D deformed TFD state $\ket{\Psi(J,\alpha(J))}$ shown in Eq.~\eqref{1D_TFD} with the entanglement temperature $T_E=1/N=0$, where $\alpha(J)$ is determined by $\frac{2\alpha}{1+\alpha^2}=\frac{(1-J)^2}{(1+J)^2}$. The phase transition point at $J_c\approx0.217$ belongs to the (1+1)D Ising universality class.}
\label{Phase_diagram_1D}
\end{figure}

\section{warm up: 1+1D deformed TFD states}\label{1_plus_1_d_TFD}
Before diving into the more complicated (2+1)D theories, we first illustrate how the deformed TFD state approach works in (1+1)D. We construct a family of $(1+1)$D deformed TFD states and study the topological phase transition in the cluster model. We focus on a special path in the phase diagram of the one-dimensional Ising cluster model that connects the fixed point SPT state and a ferromagnetic state~\cite{Wolf-Cirac-2008}. We show that the deformed TFD state approach can describe the generic (1+1)D Lorentz invariant quantum critical point, while the deformed wavefunction is limited to the (1+0)D Ising model.


The (1+1)D cluster model is defined by
\begin{equation}\label{cluster}
H_0=\sum_{i}\sigma^z_{i-1}\sigma^x_i\sigma^z_{i+1},
\end{equation}
where $\sigma^x$ and $\sigma^z$ are Pauli matrices.
It is a fixed point Hamiltonian of a non-trivial SPT state protected by a $\mathbb{Z}_2^T$ symmetry:  $\prod_i\sigma^x_iK$, where $K$ is the complex conjugate. The ground state can be expressed as:
\begin{equation}
    \ket{\psi_0}=\prod_i(1-\sigma^z_{i-1}\sigma^x_i\sigma^z_{i+1})\sum_{\pmb{\sigma}}\ket{\pmb{\sigma}},
\end{equation}
 where $\ket{\pmb{\sigma}}=\ket{\sigma_1,\cdots, \sigma_N}$.

To study the quantum phase transition between the cluster state and other states, we consider a deformed cluster ground state with a tuning parameter $J\in[0,1]$~\cite{MPS_phase_transition}:
\begin{equation}\label{RK_SPT}
|\psi(J)\rangle=\prod_i\left(1+J \sigma^z_i\sigma^z_{i+1}\right)\ket{\psi_0}.
\end{equation}
When $J=0$, $\ket{\psi(0)}=\ket{\psi_0}$ is the ground state of the cluster model $H_0$; when $J=1$, the deformation $\prod_i(1+J \sigma^z_i\sigma^z_{i+1})$ becomes a projector to the 1D ferromagnetic states. Interestingly, the deformed wavefunction can be exactly expressed as a matrix product state:
\begin{equation}\label{MPS_SPT}
|\psi(J)\rangle=\sum_{\pmb{\sigma}} \Tr(A^{[\sigma_1]}A^{[\sigma_2]}\cdots A^{[\sigma_N]})\ket{\sigma_1,\cdots, \sigma_N},
\end{equation}
where
\begin{equation}
A^{[0]}=\left(\begin{array}{cc}
   0 & 0 \\
 1 & 1
\end{array}\right),\quad
A^{[1]}=\left(\begin{array}{cc}
    1 & -(1-J)^2/(1+J)^2 \\
    0 & 0
\end{array}\right).
\end{equation}
As shown in the Appendix.~\ref{SPT_2_Ising}, the norm of $\ket{\psi(J)}$ is a partition function of a classical 1D Ising model,
\begin{equation}
\langle\psi(J)|\psi(J)\rangle\propto\sum_{\pmb{\sigma}}\exp\left[\text{arctanh}\left(\frac{2J}{1+J^2}\right)\sum_i\sigma_i\sigma_{i+1}\right].
\end{equation}
The phase diagram of the deformed SPT state $\ket{\psi(J)}$ is shown in Fig.~\ref{Phase_diagram_1D} (a). For $J\in[0,1)$, the deformed SPT state $\ket{\psi(J)}$ belongs to the paramagnetic phase. When $J=1$, the deformed SPT state $\ket{\psi(J)}$ becomes a ferromagnetic state. So, a $(1+0)$D quantum phase transition happens at $J=1$. Actually, if one considers the parent Hamiltonian of the deformed SPT state $\ket{\psi(J)}$, the phase transition point at $J=1$ is a tricritical point of the Ising cluster model with a dynamical critical exponent $z=2$~\cite{MPS_phase_transition,Skeleton_2021,pivot_2023,Crossing_2020}.

The SPT phase and phase transition can be detected by a string order parameter~\cite{String_order_2008,String_order_2012}:
\begin{equation}\label{string_order_para}
    O=\sqrt{\lim_{|i-k|\rightarrow\infty}\left[{\bra{\psi(J)} S_{ik}\ket{\psi(J)}}/{\langle \psi(J)|\psi(J)\rangle}\right]},
\end{equation}
where
\begin{equation}\label{string_operator}
 S_{ik}=\sigma_i^z\sigma_{i+1}^y\left(\prod_{j=i+2}^{k-2}\sigma^x_j\right)\sigma_{k-1}^y\sigma_k^z,
\end{equation}
is the SPT string operator.
Using MPS representation in Eq.~\eqref{MPS_SPT}, the string order parameter $O$ and the correlation length $\xi$ of the deformed SPT state can be exactly calculated~\cite{Crossing_2020,xu_2023}:
\begin{equation}
    O=\frac{1-J^2}{1+J^2}\sim(1-J), \quad \xi=-1/\log(\frac{1+J^2}{2J})\sim(1-J)^{-2},\notag
\end{equation}
from which we find the critical exponents $\beta_1 = 1$ and $\nu_1=2$ according to $O\sim (1-J)^{\beta_1}$ and $\xi\sim (1-J)^{-\nu_1}$ respectively.

To be able to describe a generic $(1+1)$D quantum critical point, we consider a two-parameter family of deformed TFD states:
\begin{align}\label{1D_TFD}
|\Psi(J,\alpha)\rangle&=\sum_{\pmb{\sigma}}\left[\prod_i\left(1+J \sigma^z_i\sigma^z_{i+1}\right)\left(1-\alpha \sigma^z_{i-1}\sigma^x_i\sigma^z_{i+1}\right)
\ket{\pmb{\sigma}}_P\right]\notag\\&\otimes\ket{\pmb{\sigma}}_F.
\end{align}
When $\alpha=1$, the physical and the fictitious degrees of freedom of the TFD state are disentangled:
\begin{equation}\label{distentangle_SPT}
    |\Psi(J,1)\rangle\propto|\psi(J)\rangle_P\otimes |\psi(0)\rangle_F,
\end{equation} 
where $|\psi(J)\rangle$ is the deformed wavefunction defined in Eq.~\eqref{RK_SPT},
and the norm of the TFD state $\langle\Psi(J,1)|\Psi(J,1)\rangle\propto\langle\psi(J)|\psi(J)\rangle$ is still a partition function of a 1D classical Ising model. However, if $\alpha<1$, the physical and auxiliary degrees of freedom are entangled with each other, which is crucial for obtaining the $(1+1)$D quantum critical point.  Moreover, the deformed TFD state in Eq.~\eqref{1D_TFD} can be exactly expressed as an MPS, because the deformation $\prod_i(1+J\sigma_i^z\sigma_{i+1}^z)$ and the projector $\prod_i(1-\alpha\sigma_{i-1}^z\sigma_i^x\sigma_{i+1}^z)$ are constant depth non-unitary circuits, which are equivalent to matrix product operators and can be reshaped as MPS.

According to the general framework, we consider a reduced density matrix of the TFD state by tracing the fictitious degrees of freedom:
\begin{align}\label{rho_TFD_1D}
\rho&=\Tr_F\ket{\Psi(J,\alpha)}\bra{\Psi(J,\alpha)}.
\end{align}
A partition function can be obtained from $\rho^N$ with an entanglement temperature $T_E=1/N$:
\begin{equation}\label{2D_classical_Ising}
\mathcal{Z}_N=\Tr(\rho^N)\propto\sum_{\pmb{\sigma}^{[1]}\cdots\pmb{\sigma}^{[N]}}\exp\left(\sum_{i,\tau}K_x\sigma^{[\tau]}_i\sigma^{[\tau]}_{i+1}+K_\tau\sigma^{[\tau]}_i\sigma^{[\tau+1]}_{i}\right),
\end{equation}
which is nothing but the partition function of a 2D  classical Ising model with anisotropic coupling strength:
\begin{align}
    &K_{\tau}=\frac{1}{2}\log\frac{1+\alpha^2}{2\alpha},\quad K_{x}=2\text{tanh}^{-1}J.
\end{align}
We can impose the relation $K_x=K_{\tau}$ such that $\alpha$ is a function of $J$ to reduce the number of parameters, and the partition function becomes an isotropic 2D classical Ising model. So we know that phase diagram of the TFD state $\ket{\Psi(J)}$ at $T_E=1/N=0$ shown in Fig.~\ref{Phase_diagram_1D} (b). There is a continuous phase transition at $J_c=1+\sqrt{2}-\sqrt{2(1+\sqrt{2})}\approx0.217$ (determined by $K_x=K_{\tau}=\log(1+\sqrt{2})/2$~\cite{Wu_Potts,Baxter_exact}) described by the (1+1)D Ising CFT with a dynamical critical exponent $z=1$~\cite{CFT_book}, in contrast to the dynamical critical exponent $z=2$ obtained from the deformed wavefunction $\ket{\psi(J)}$ in Eq.~\eqref{RK_SPT}. Furthermore, we can also evaluate the expectation value of the string operator $S_{ik}$ in Eq.~\eqref{string_operator} using the TFD state at $T_E=0$:
 \begin{equation}
    O=\sqrt{\lim_{|i-k|\rightarrow\infty}\left[\Tr\left( S_{ik}\rho^N(J)\right)/\Tr \rho^N(J)\right]}.
\end{equation}
It turns out that the string order parameter is equivalent to the disorder parameter evaluated using the partition function of a 2D classical Ising model, see Appendix.~\ref{SPT_2_Ising}. So we know when $T_E=1/N=0$, $O\sim(J-J_c)^{\beta_2}$ and $\xi\sim(J-J_c)^{-\nu_2}$ with the critical exponents $\beta_2=1/8$ and $\nu_2=1$~\cite{CFT_book}. Therefore, we obtain a quantum phase transition with emergent Lorentz invariance from the deformed (1+1)D TFD state~\eqref{1D_TFD}.

\begin{figure}[tbp]
\centering
\includegraphics[width=\columnwidth]{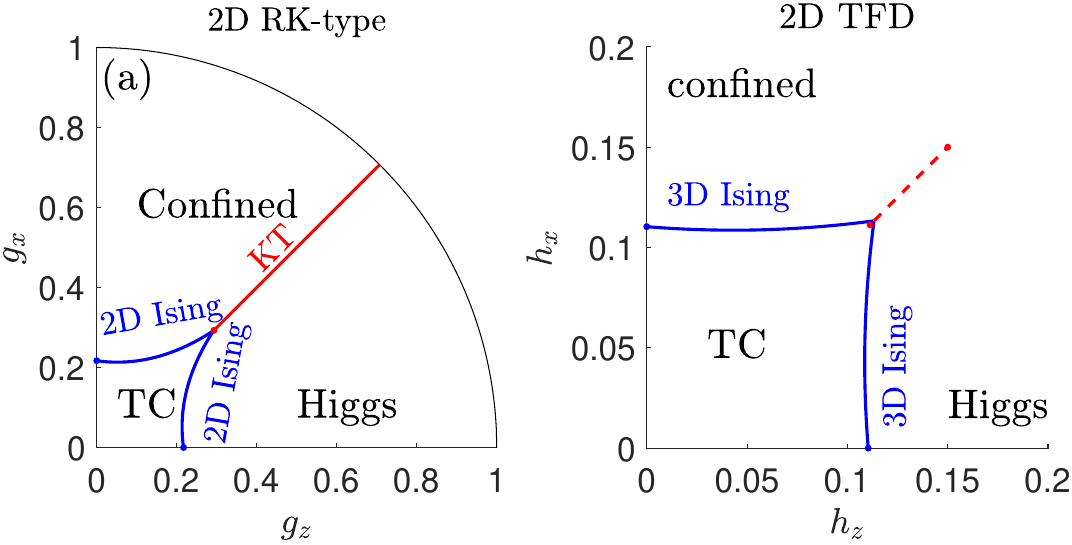}
\caption{(a) The phase diagram of the deformed toric code state $\ket{\psi(g_x,g_z)}$ in Eq.~\eqref{deformed_TC}, TC is the gapped toric code phase with the  $\mathbb{Z}_2$ topological order, the confined (Higgs) phase is a trivial gapped phase. The continuous phase transitions represented by the blue lines are described by 2D Ising CFT. The red line is a gapless KT phase described by a family of 2D CFTs with a central charge $c=1$. (b) The phase diagram of the deformed TFD state $\ket{\Psi(h_x,a(h_z),h_z,b(h_x))}$ in Eq.~\eqref{2D_TFD} at the entanglement temperature $T_E=1/N=0$, where $a$ ($b$) is determined by $h_z$ ($h_x$). The continuous phase transition represented by the blue lines is described by the (2+1)D Ising CFT. The confined phase and Higgs phase are the same phase because there is a first-order transition line (red dash line)  terminating at finite $h_x$ and $h_z$.}
\label{2D_phase_diagram}
\end{figure}

\section{(2+1)D deformed TFD states: Toric code model}\label{2_plus_1_d_TFD}
In this section, we construct a $(2+1)$D deformed TFD state for topological phase transitions between the toric code phase and a trivial paramagnetic phase.
The topological phase transitions of the toric code ground state deformed by string tensions has been studied~\cite{Simon_Trebst_2007,zhu_gapless_2019}, where the deformed toric code states can be exactly expressed in terms of PEPS with a constant bond dimension 2. The continuous phase transitions from the toric code phase to a trivial phases are described by the $(2+0)$D Ising CFT. However, the generic topological phase transition from the toric code phase to the trivial phase should be characterized by a $(2+1)$D Ising CFT. We show how to obtain this generic topological phase transition from the deformed TFD state approach.

 Consider a toric code model defined on a square lattice~\cite{kitaev_toric_code_2003}:
\begin{eqnarray}
H_{\text{TC}} &=&-\sum_{v}A_{v}-\sum_{p}B_{p}, \\
A_{v} &=&-\sum_{v}\prod_{\langle ij\rangle \in v}\sigma _{ij}^{x},\quad
B_{p}=-\sum_{p}\prod_{\langle ij\rangle \in p}\sigma _{ij}^{z},  \notag
\end{eqnarray}%
where the vertex and plaquette terms $A_{v}$ and $B_{p}$ involve four Pauli
matrices located on the bonds between sites $i$ and $j$. A ground state can be
constructed using the projectors:
\begin{equation}\label{TC_GS}
|\psi _0\rangle =\prod_{v} \left(1+A_{v}\right) \prod_{p}\left(
1+B_{p}\right) \sum_{\pmb{\sigma}}\ket{\pmb{\sigma}}.
\end{equation}%
where $\ket{{\pmb{\sigma}}}$ is a set of complete and orthonormal basis.


In order to study topological quantum phase transitions out of the
topological phase, a deformed toric code state has been proposed~\cite{haegeman_shadows_2015,GaugingTNS_2015,zhu_gapless_2019}:
\begin{equation}\label{deformed_TC}
\ket{\psi(g_x,g_z)}=\prod_{\langle ij\rangle}(1+g_x\sigma^x_{ij}+g_z\sigma^z_{ij})|\psi_0\rangle,
\end{equation}%
where $g_x$ and $g_z$ are tuning parameters satisfying $g_x,g_z>0$ and $g_x^2+g_z^2=g^2\leq1$. By introducing two sets of auxiliary degrees of freedom $\pmb{s}=\{s_i\}$ and $\pmb{t}=\{t_i\}$ on vertices of the square lattice,  it can be found that the norm of the deformed toric code state $\langle\psi(g_x,g_z)|\psi(g_x,g_z)\rangle$ is proportional to the partition function of the 2D classical Ashkin-Teller (AT) model~\cite{zhu_gapless_2019}:
\begin{equation}\label{AT_model}
\mathcal{Z}_{\text{AT}}= \sum_{\pmb{s},\pmb{t}}\prod_{\langle ij\rangle}\exp\left[J_2(s_is_j+t_it_j)+J_4s_is_jt_it_j\right],
\end{equation}
where
\begin{equation}
    J_2=\frac{1}{4}\log\frac{1+g^2+2g_z}{1+g^2-2g_z}, J_4=\frac{1}{4}\log\frac{1+g^4+2g_x^2-2g_z^2}{2g_x^2-2g_z^2+2g^2}\notag.
\end{equation}
So the phase diagram of the deformed toric code state $\ket{\psi(g_x,g_z)}$ can be obtained from that of the AT model, see Fig.~\ref{2D_phase_diagram} (a). There is a gapped toric code phase, and a gapped Higgs (confined) phase in which the charges condense (are confined). The continuous phase transitions from the toric code phase to Higgs or confined phase belong the 2D Ising universality class. Along the self-dual line $h_x=h_z$, there is a phase transition from the toric code phase to a gapless Kosterlitz-Thouless (KT) phase through a KT transition, and the gapless KT phase seperates the Higgs and confined phases.

\begin{figure}[h]
\centering
\includegraphics[width=\columnwidth]{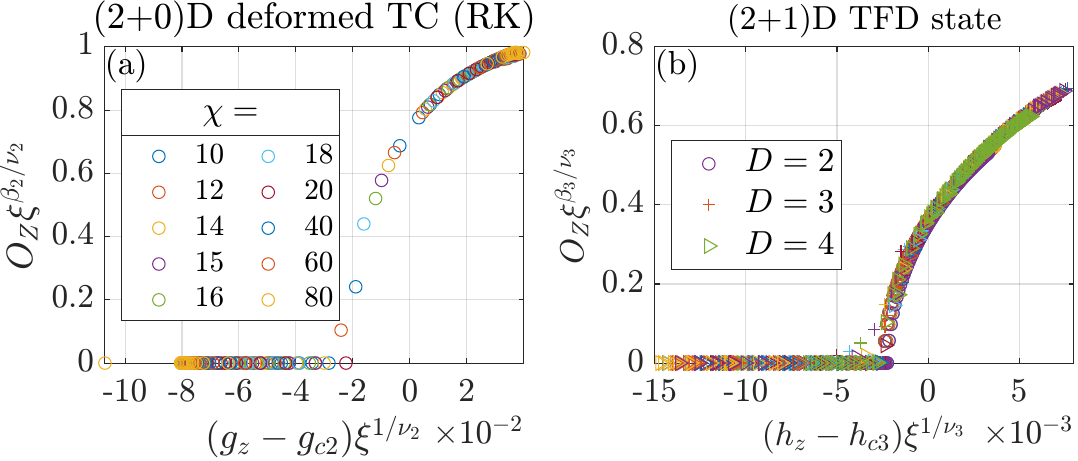}
\caption{Data collapse for the string order parameter $O_Z$, where $\xi$ is the correlation length. (a)  Results for the 2D RK-type deformed toric code state shown in Eq.~\eqref{deformed_TC} with $g_x=0$, where $g_{c2}=1+\sqrt{2}+\sqrt{2(1+\sqrt{2})}$, $\nu_2={1}$, $\beta_2=1/8$, $\chi$ is the bond dimension of the  boundary infinite matrix product states (iMPS) for contracting the 2D partition function in Eq.~\eqref{AT_model}. (b) Results for the deformed 2D TFD state in Eq.~\eqref{2D_TFD} with $h_x=0$, where $h_{c3}\approx0.110376$, $\nu_3\approx0.629 971$,$\beta_3\approx0.326 419$, $D$ is the bond dimension of the boundary infinite PEPS for contracting the 3D partition function in Eq.~\eqref{3D_partition_function}, the bond dimensions $\chi$ of the iMPS for contracting infinite PEPS ranges from 10 to 100.}
\label{data_collapse}
\end{figure}

A string order parameter can be introduced to describe the topological phase transition along $g_z$ axis ($g_x=0$):
\begin{equation}
    O_Z=\left|\lim_{|L|\rightarrow\infty}\frac{\bra{\psi(0,g_z)}\prod_{\langle ij\rangle\in L}\sigma^z_{ij}\ket{\psi(0,g_z)}}{\langle\psi(0,g_z)|\psi(0,g_z)\rangle}\right|^{1/2},
\end{equation}
where $L$ is a string along the sqaure lattice and $|L|$ is the distance between two ends of $L$. Since the AT model in Eq.~\eqref{AT_model} reduces to the Ising model when $g_x=0$, it can be found that the string order parameter can be transformed to a local Ising order parameter. So we have $O_Z\sim(g_z-g_{c2})^{\beta_2}$ and the correlation length $\xi\sim(g_z-g_{c2})^{-\nu_2}$, where the critical point is $g_{c2}=1+\sqrt{2}-\sqrt{2(1+\sqrt{2})}$~\cite{Wu_Potts,Baxter_exact}, and the critical exponents $\beta_2=1/8$ and $\nu_2=1$ from the 2D Ising universality class~\cite{CFT_book}. In Fig.~\ref{data_collapse} (a), we numerically check the critical exponents use the data collapse based on the finite entanglement scaling~\cite{Bram_2019}, which matches perfectly with the 2D Ising universality class.

However, we know that the phase diagram of the deformed toric code state $\ket{\psi(g_x,g_z)}$ phase is qualitatively different from the phase diagram of the toric code  Hamiltonian in a parallel magnetic field~\cite{youjin2012,Vidal2009,TC_in_xyz_field_2011}, where the continuous phase transitions from the toric code phase to the Higgs (confined) phase belongs to the (2+1)D Ising universality class, and the Higgs phase and confined phase can be smoothly connected are the same phase.

In order to obtain the generic topological phase transitions of the toric code model, we consider the following deformed TFD state:
\begin{align}\label{2D_TFD}
&\ket{\Psi(h_x,a,h_z,b)}=\sum_{\pmb{\sigma}}\left[\mathbb{T}_{xA}(h_x,a)\mathbb{T}_{zB}(h_z,b)\ket{\pmb{\sigma}}_P\right]\otimes\ket{\pmb{\sigma}}_F,
\end{align}
where the parameters $h_{x},h_{z},a,b\in \lbrack 0,1]$ and
\begin{align}
    \mathbb{T}_{xA}(h_x,a)&=\prod_{\langle ij \rangle}(1+h_x\sigma^x_{ij})\prod_{v}(1+a A_v),\notag\\
    \mathbb{T}_{zB}(h_z,b)&=\prod_{\langle ij \rangle}(1+h_z\sigma^z_{ij})\prod_{p}(1+b B_p).
\end{align}
A reduced density matrix can be obtained by tracing out the fictitious degrees of freedom:
\begin{align}
\rho =\text{tr}_{F}|\Psi\left( h_{x},a,h_{z},b\right) \rangle
\langle \Psi\left( h_{x},a,h_{z},b\right) |=\mathbb{T}\mathbb{T}^{\dagger},
\end{align}%
where $\mathbb{T}=\mathbb{T}_{xA}\mathbb{T}_{zB}$.

\begin{figure}[tbp]
\centering
\includegraphics[width=0.8\columnwidth]{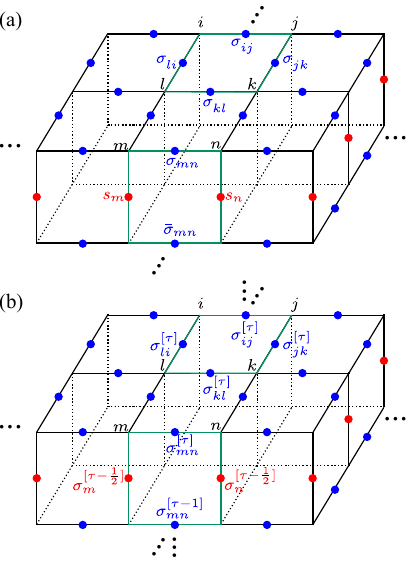}
\caption{(a) The distribution of degrees of freedom of $\mathbb{T}_{xA}\mathbb{T}_{zB}$, the blue and red dots are the physical and auxiliary degrees of freedom. (b) Distribution of degrees of freedom in one temporal layer of the partition function $\mathcal{Z}_N$, which are obtained by relabeling the degrees of freedom in $\mathbb{T}_{xA}\mathbb{T}_{zB}$.}
\label{TN}
\end{figure}

By introducing auxillary degress of freedom $\pmb{s}=\{s_i\}$, $\mathbbm{T}_{xA}$ and $\mathbb{T}_{zB}$ can be expressed as:
\begin{align}\label{PEPO_T_xA_zB}
\mathbb{T}_{xA} &\propto \sum_{\pmb{s}\pmb{\sigma}\bar{\pmb{\sigma}}}\prod_{i}e^ %
 {-\log (a)s_{i}/2}\prod_{\langle ij\rangle}e^{-\log (h_{x})s_{i}s_{j}\sigma
_{ij}\bar{\sigma}_{ij}/2} |\pmb{\sigma}\rangle \langle \bar{\pmb{\sigma}}%
|,  \notag \\
\mathbb{T}_{zB} &\propto \sum_{\pmb{\sigma}}\prod_{\langle ij\rangle}e^{\tanh ^{-1}(h_{z})\sigma _{ij}}\prod_{p}e^{\tanh ^{-1}(b)\prod_{\langle ij\rangle \in p}\sigma
_{ij}} |\pmb{\sigma}\rangle \langle \pmb{\sigma}|,
\end{align}%
where the distribution of the degrees of freedom is shown in Fig.~\ref{TN} (a).
When $a=b=1$, we obtained a partition function of the deformed TFD state at $T_E=1$:
\begin{align}
   \label{2_decouple_Ising} \mathcal{Z}_1&=\Tr\rho\propto\sum_{\pmb{\sigma}\pmb{s}}\prod_p(1+\prod_{\langle ij\rangle\in p}\sigma_{ij})e^{\sum_{\langle ij\rangle}\left(\tilde{J}_x s_i s_j+\tilde{J}_z\sigma_{ij}\right)}\notag\\
    &\propto\sum_{\pmb{s}\pmb{u}}\exp\left[\sum_{\langle ij\rangle}\left(\tilde{J}_x s_i s_j+\tilde{J}_zu_ju_j\right)\right].
\end{align}
where
\begin{equation}\label{JxJz}
  \tilde{J}_x=\frac{1}{2}\log\frac{1+h_x^2}{2h_x},\quad \tilde{J}_z=2\tanh^{-1}h_z,
\end{equation}
and a new $\mathbb{Z}_2$ spin $u_i$ on a site $i$ of the lattice is defined via $\sigma_{ij}=u_iu_j$. Compared to the norm of the deformed toric code state $\ket{\psi(g_x,g_z)}$ in Eq.~\eqref{AT_model}, which is the partition function of the AT model,  we know the deformed toric code state in Eq.~\eqref{deformed_TC} is not included in the deformed TFD state because the partition function $\mathcal{Z}_1$ in Eq.~\eqref{2_decouple_Ising} is a two-decoupled Ising model not the AT model. This is reasonable because the 3D AT model is not the partition function for the generic topological phase transition of the toric code model~\cite{Machine_learning_TC_SD_2023}. The deformed TFD state $\ket{\Psi}$ is a more natural starting point than the deformed toric code state $\ket{\psi}$ to reach the generic topological phase transitions of the toric code model.
Only if $h_x=g_x=0$ or $h_z=g_z=0$, the deformed TC state $\ket{\psi}$ is included in the deformed TFD state $\ket{\Psi}$
\begin{align}\label{disentangle_topo}
|\Psi(h_x,1,0,1)\rangle&\propto\ket{\psi(g_x,0)}_P\otimes\ket{\psi(0,0)}_F,\notag\\
|\Psi(0,1,h_z,1)\rangle&\propto\ket{\psi(0,0)}_P\otimes\ket{\psi(0,g_z)}_F.
\end{align}
Interestingly, when $h_x=0,h_z=1,b=0$, the deformed TFD state becomes
\begin{equation}
  |\Psi(0,a,1,0)\rangle\propto\left[\prod_{v}(1+aA_v)\ket{00\cdots0}_P\right]\otimes\ket{00\cdots0}_F,
\end{equation}
which can be prepared on a quantum computer in terms of a quantum circuit~\cite{Sun_2023}.

In order to reduce the entanglement temperature $T_E$, we consider $\rho^N$ and the partition function $\mathcal{Z}_N=\Tr\rho^N$ with $T_E=1/N$. Using Eq.~\eqref{PEPO_T_xA_zB}, we can construct the explicit form of the partition
function (ignoring the unimportant coefficients):
\begin{eqnarray*}
&&\mathcal{Z}_{N} =\sum_{\substack{ \pmb{\sigma}^{[1]},\pmb{\sigma}^{[\frac{1}{%
2}]},\cdots ,  \\ \pmb{\sigma}^{[N+\frac{1}{2}]},\pmb{\sigma}^{[N]}}}%
\prod_{\tau =1}^{N}\exp \left[ \tilde{J}_{A}\sum_{i}\sigma _{i}^{[\tau +\frac{1%
}{2}]}+\tilde{J}_{z}\sum_{\langle ij\rangle }\sigma _{ij}^{[\tau ]}\right. \\
&&\left. +\tilde{J}_{x}\sum_{\langle ij\rangle }\sigma _{i}^{[\tau +\frac{1}{2}%
]}\sigma _{j}^{[\tau +\frac{1}{2}]}\sigma _{ij}^{[\tau ]}\sigma _{ij}^{[\tau
+1]}+\tilde{J}_{B}\sum_{p}\prod_{\langle ij\rangle \in p}\sigma _{ij}^{[\tau ]}%
\right] ,
\end{eqnarray*}%
where we have substituted the labels $\sigma _{ij}^{[\tau ]},\sigma
_{ij}^{[\tau +1]},\sigma _{i}^{[\tau +\frac{1}{2}]}$ for the labels $\sigma
_{ij},\bar{\sigma}_{ij},s_{i}$, as shown in Fig.~\ref{TN} (b), and
\begin{align}
    \tilde{J}_A &=\frac{1}{2}\log\frac{1+a^2}{2a},\quad \tilde{J}_B=2 \tanh^{-1} b,
\end{align}
$\tilde{J}_x$ and $\tilde{J}_z$ are defined in Eq.~\eqref{JxJz}.
When the isotropic inter-layer and intra-layer interactions are required, we
can impose the following relations to the parameters: $\tilde{J}_{A} =\tilde{J}_{z}, \tilde{J}_{x} =\tilde{J}_{B}$, and the number of the deformed parameters are reduced to $2$ and the
partition function can be simplified as
\begin{equation}\label{isotropic_GH}
\mathcal{Z}_{N}=\sum_{\pmb{\sigma}}\exp \left[ \tilde{J}_z\sum_{\langle \pmb{i}\pmb{j}%
\rangle }\sigma _{\pmb{i}\pmb{j}}+\tilde{J}_x\sum_{\pmb{p}}\prod_{\langle \pmb{i}%
\pmb{j}\rangle \in \pmb{p}}\sigma _{\pmb{i}\pmb{j}}\right] ,
\end{equation}%
where $\pmb{i}=(i,\tau )$ denotes the cubic lattice sites, $\pmb{p}$ is a
plaquette of the cubic lattice, and $\pmb{\sigma}$ stands for the
configurations of all spins $\sigma _{\pmb{ij}}$ on the bonds of the cubic
lattice. Actually this partition
function is nothing but the 3D classical $\mathbb{Z}_2$ gauge Higgs model~\cite%
{Kogut_LGH,Fradkin_1979} which can also be derived from the quantum-classical mapping of a 2D toric code Hamiltonian in a parallel magnetic field via a Trotter decomposition of an imaginary time evolution~\cite{Kogut_Susskind,Gauge_Higgs_2010,Map_TC_Z_2_GH}.
From the known phase diagram of the 3D classical $\mathbb{Z}_2$ gauge Higgs model~\cite{Gauge_Higgs_1980,Gauge_Higgs_2010,gauge_Higgs_2021,Gauge_higgs_2022}, we can plot the phase diagram of the TFD state at the entanglement temperature $T_E=0$, as shown in Fig.~\ref{2D_phase_diagram} (b), which contains the generic topological phase transitions of the toric code model.

Along the $h_{z}$ axis ($h_x=0$), $J_{x}\rightarrow\infty$, the
partition function shown in Eq.~\eqref{isotropic_GH} reduces to
\begin{equation}\label{3D_partition_function}
\mathcal{Z}_{N}=\sum_{\pmb{\sigma}}\prod_{\pmb{p}}\left( 1+\prod_{{\langle %
\pmb{i}\pmb{j}\rangle }\in \pmb{p}}\sigma _{\pmb{i}\pmb{j}}\right) \exp \left[
J_z\sum_{\langle \pmb{i}\pmb{j}\rangle }\sigma _{\pmb{i}\pmb{j}}\right].
\end{equation}
By introducing new spin variables $u _{\pmb{i}}$ at the cubic lattice sites
$\pmb{i}$, $\sigma _{\pmb{i}\pmb{j}}$ on a bond between the
sites $\pmb{i}$ and $\pmb{j}$ can be represented by the new variables\cite{Simon_Trebst_2007}: $\sigma _{\pmb{i}\pmb{j}}=u _{\pmb{i}}u _{\pmb{j}}$, and above
 the partition function is transformed to that of 3D classical Ising model on a cubic lattice $\mathcal{Z}_{N}=\sum_{\mu }\exp \left[ J_z\sum_{\langle \pmb{i}\pmb{j}\rangle
}u _{\pmb{i}}u _{\pmb{j}}\right]$. So we know that when $h_x=0$ there is a continuous phase transition at the critical point is $h_z=h_{c3}\approx0.110376$~\cite{talapov_1996_magnetization} from the toric code phase to the trivial phase.
We can also evaluate the string order parameter using the TFD state at $T_E=1/N$,
\begin{equation}
    O_Z=\left|\lim_{|L|\rightarrow\infty}\left[\Tr\left(\rho^N\prod_{\langle ij\rangle\in L}\sigma^z_{ij}\right)/\Tr\left(\rho^N\right)\right]\right|^{1/2}.
\end{equation}
When $T_E=0$, the string order parameter can be transformed to a local order parameter of the 3D classical Ising model. So we have $O_Z\sim(h_z-h_{c3})^{\beta_3}$ and the correlation length $\xi\sim(h_z-h_{c3})^{-\nu_3}$, where the critical exponents $\beta_3\approx0.326419$ and $\nu_3\approx0.629971$ are from 3D Ising CFT~\cite{Conformal_bootstrap_2016}. The local order parameter of the 3D classical Ising model can be obtained using a tensor network method~\cite{TNS_3D_classical_2018}. In Fig.~\ref{data_collapse} (b), we numerically check the critical exponents using the finite-entanglement scaling~\cite{Corboz-prx-2018,Lauchli-prx-2018,Bram_2022}, which matches perfectly with the 3D Ising universality class.

\section{Conclusion and Outlook}\label{sec_summary}
We propose a framework of deformed TFD states which bridges the RK-type and generic quantum phase transitions. From the quantum-classical mapping of the deformed TFD states, we can construct the missing temporal direction in the quantum-classical mapping of the RK-type deformed wavefunctions. These ideas are illustrated by a family of 1D deformed TFD states for a topological phase transition between a SPT phase and a ferromagnetic phase, and a family of 2D deformed TFD states for topological phase transitions between a toric code phase and a paramagnetic phase. From our results, we can understand the missing ingredients in the RK-type deformed wavefunctions: (i) entanglement between the degrees of freedom in $\mathscr{H}_P$ and $\mathscr{H}_F$; (ii) approaching generic quantum critical points by reducing the entanglement temperature, which effectively increasing the bond dimensions such that we get rid of the constant bond dimension PEPS manifold.

When the physical and fictitious degrees of freedom are disentangled, the deformed TFD states naturally reduce to the RK-type deformed wavefunctions, as shown in Eqs.~\eqref{disentangle}, \eqref{distentangle_SPT} and ~\eqref{disentangle_topo}, so the deformed TFD states can have the SPT order or the topological order. However, the SPT order and topological order are lost in the deformed TFD states when physical and fictitious degrees of freedom are entangled because 1D SPT or 2D topological order do not exit at any finite temperature~\cite{Hastings_2011,SPT_at_finite_T_2017}. However, the SPT order or topological order can be recovered when $N\rightarrow\infty$ and $T_E=0$. It would be interesting to study TFD states satisfying: (i) the physical and fictitious degrees of freedom are entangled; (ii) and having SPT order or topological order; (iii) can be exactly expressed as finite constant bond dimension TNS.

There are some possible applications of the deformed TFD states. First, the deformed TFD state with a finite $N$ can be used as a variational ansatz for quantum systems at a finite temperature. Second, although variational 2D infinite PEPS have been used to numerically calculate the static critical exponents at the Lorentz invariant quantum
critical points~\cite{Corboz-prx-2018,Lauchli-prx-2018,Schuch-2020}, the dynamical critical exponent $z$ is not easy to be extracted; this might be improved by taking the TFD states into consideration. Third, since the deformed TFD states can be expressed exactly as TNS, which can be transformed into quantum circuits exactly in 1D and possibly in 2D~\cite{Prepare_PEPS_quamtum_computer}, it is possible to realize Lorentz invariant (topological) quantum phase transitions using deformed TFD states on a quantum computer instead of the RK-type~\cite{TFD_quantum_circuit_PRL_2019,TFD_on_Quantum_computer_2020,TFD_on_Quantum_computer_2021,Crossing_2020}.

\section*{Acknowledgements}
The authors appreciate Bram Vanhecke and Laurens Vanderstraeten for providing the optimized boundary infinite PEPS tensors of the 3D classical Ising model. The authors are grateful to Guo-Yi Zhu and Qi Zhang for the stimulating discussions. W.-T. Xu acknowledges support from the Munich Quantum Valley, which is supported by the Bavarian
state government with funds from the Hightech Agenda
Bayern Plus. R.-Z. Huang is supported by a postdoctoral fellowship from the Special Research Fund (BOF) of Ghent University. The research is supported by the National Key R\&D Program of China
(2023YFA1406400).

\section*{Data availability} Data, data analysis, and simulation codes are available upon reasonable request on Zenodo~\cite{zenodo}.
\appendix

\section{Quantum-classical mapping for the (1+1)D case}\label{SPT_2_Ising}

In this Appendix, we map the deformed 1D SPT state and the deformed 1D TFD state to the 1D and 2D classical Ising models, separately. It is known that the cluster model $H_0$ in Eq.~\eqref{cluster} is related to the fixed point Hamiltonian $H_1=-\sum_i \sigma^x_i$ of a paramagnetic phase by a unitary transformation~\cite{RK_SPT_2015}: $H_0=UH_1U^{\dagger}$, where
\begin{equation}
U=\prod_iCZ_{i,i+1}\prod_i\sigma^z_i,
\end{equation}
and $CZ_{i,i+1}=(1+\sigma^z_i+\sigma^z_{i+1}-\sigma^z_{i}\sigma^z_{i+1})/2$ is the control-Z gate. So the ground state of the cluster model $H_0$ can also be expressed as
\begin{align}\label{clsuter_state}
  |\psi_0\rangle&=U\ket{++\cdots+}\propto\prod_iCZ_{i,i+1}\prod_i\sigma^z_i\sum_{\pmb{\sigma}}\ket{\pmb{\sigma}},
\end{align}
where $\ket{+}$ is an eigenstate of $\sigma^x$ with an eigenvalue $1$. From Eq.~\eqref{clsuter_state}, we can derive that the norm of the deformed SPT state is a partition function of a 1D classical Ising model:
\begin{align}
\braket{\psi(J)}{\psi(J)}=&\sum_{\pmb{\sigma}^{\prime},\pmb{\sigma}}\bra{\pmb{\sigma}^{\prime}}U^{\dagger}(1+J^2+2J\sigma_i^z\sigma_{i+1}^z)U\ket{\pmb{\sigma}}\notag\\
=&\sum_{\pmb{\sigma}^{\prime},\pmb{\sigma}}\bra{\pmb{\sigma}^{\prime}}(1+J^2+2J\sigma_i^z\sigma_{i+1}^z)\ket{\pmb{\sigma}}\notag\\
\propto&\sum_{\pmb{\sigma}}\exp\left[\text{arctanh}\left(\frac{2J}{1+J^2}\right)\sum_i\sigma_i\sigma_{i+1}\right].
\end{align}

Next, let us derive the partition function from the deformed TFD state. We can obtain the reduced TFD density matrix by tracing ficticious degrees of freedom:
\begin{align}
\rho&=\Tr_F\ket{\Psi(J,\alpha)}\bra{\Psi(J,\alpha)}=\prod_i(1+J \sigma^z_i\sigma^z_{i+1})\notag\\
&\times\prod_i(1+\alpha^2-2\alpha\sigma^z_{i-1}\sigma^x_i\sigma^z_{i+1})\prod_i(1+J\sigma^z_i\sigma^z_{i+1}),
\end{align}
which can be simplified using the unitary transformation $U$:
\begin{equation}
\rho^{\prime}=U\rho U^{\dagger}=\prod_i(1+J\sigma^z_i\sigma^z_{i+1})(1+\alpha^2+2\alpha \sigma^x_i)(1+J \sigma^z_i\sigma^z_{i+1}),\notag
\end{equation}
such that we have $\mathcal{Z}_N=\Tr(\rho^N)=\Tr[(\rho^{\prime})^N]$. Because
\begin{align}
&\bra{\sigma^{[\tau]}}(1+\alpha^2+2\alpha\sigma^x)\ket{\sigma^{[\tau+1]}}\propto \exp(K_{\tau}\sigma^{[\tau]}\sigma^{[\tau+1]}),\notag\\
&\bra{\sigma_i^{[\tau]}\sigma_{i+1}^{[\tau]}}(1+J^2+2J\sigma^z_i\sigma^z_{i+1})\ket{\sigma_i^{[\tau]}\sigma^{[\tau]}_{i+1}}\propto e^{K_{x}\sigma^{[\tau]}_i\sigma^{[\tau]}_{i+1}},\notag
\end{align}
we can insert the operators $\sum_{\pmb{\sigma}}\ket{\pmb{\sigma}^{[\tau]}}\bra{\pmb{\sigma}^{[\tau]}}$ with $\tau=1,2,\cdots,N$ into $(\rho^{\prime})^N$ and derive the partition function in Eq.~\eqref{2D_classical_Ising}. Moreover, using the unitary transformation $U$, the SPT string operator in Eq.~\eqref{string_operator} is simplified to the disorder operator of the Ising model~\cite{fradkin2017disorder},
\begin{equation}
US_{ik}U^{\dagger}=\prod_{j=i+1}^{k-1}\sigma^x_j.
\end{equation}
Under the Kramers-Wainner duality of the 2D classical or $(1+1)$D quantum Ising model, the disorder parameter is dual to the local order parameter, which has a critical exponent $\beta=1/8$~\cite{CFT_book}. Therefore we know the critical exponent of the SPT string order parameter evaluated using the TFD state is $1/8$.

 \bibliography{3d_model_ref.bib}
\end{document}